\documentclass[amssymb,aps,twocolumn,showpacs]{revtex4}
\usepackage[]{graphicx}
\usepackage[]{amsmath}
\usepackage{hyperref}

\def\Tr{\mathrm{Tr}}
\def\II{1\!\mathrm{l}}

\def\cC{\mathcal{C}}

\def\cH{\mathcal{H}}

\def\sn{\star^{(n)}}

\def\Eq#1{Eq.~\eqref{eq:#1}}

\begin{document}

\title{Belief propagation algorithm for computing correlation functions in finite-temperature quantum many-body systems on loopy graphs}
\author{David Poulin$^1$}
\author{Ersen Bilgin$^2$}
\affiliation{$^1$Center for the Physics of Information, $^2$Physics Department, Caltech, Pasadena, CA 91125}

\date{\today}

\begin{abstract}
Belief propagation --- a powerful heuristic method to solve inference problems involving a large number of random variables  --- was recently generalized to quantum theory. Like its classical counterpart, this algorithm is exact on trees when the appropriate independence conditions are met and is expected to provide reliable approximations when operated on loopy graphs. In this paper, we benchmark the performances of loopy quantum belief propagation (QBP) in the context of finite-temperature quantum many-body physics. Our results indicate that QBP provides reliable estimates of the high-temperature correlation function when the typical loop size in the graph is large. As such, it is suitable e.g. for the study of quantum spin glasses on Bethe lattices and the decoding of sparse quantum error correction codes. 
\end{abstract}


\maketitle

\section{Introduction}

Belief propagation is a powerful algorithm designed to solve inference problems involving a large number of random variables. It operates on graphical models, where variables are located at the vertices of a graph and edges encode dependence relations between the variables. The algorithm is exact when the underlying graph is a tree, but most importantly, it performs remarkably well in circumstances where it is not proven to converge to the exact solutions, i.e. when the graphical model contains loops. It is also highly parallelizable in the sense that each random variable can be associated with a different processor, and messages are exchanged between processors that are joined by an edge \cite{AM00a,Yed01a,Mac03a,MM07a}. 

These features have made belief propagation an important tool in numerous scientific and technological fields ranging from information theory to image recognition, and from artificial intelligence to statistical physics. Indeed, it is one of the most powerful heuristic algorithms to solve problems such as decoding of low-density and turbo error correction codes \cite{Gal63a,BGT93a,RU05a}, determining the phase diagram of quenched disordered systems \cite{MP01a,MM07a}, and random satisfiability problems \cite{MZK+99a,MPZ02a}.

Recently, belief propagation and graphical models were generalized to the quantum setting \cite{LP07a,Has07b}.  In this article, we characterize the performances of QBP when used as a heuristic algorithm to solve inference problems --- e.g. compute correlation functions --- in the context of finite-temperature quantum many-body physics. 

\section{Graphical models}

We consider quantum graphical models $(G,\rho)$ that consist of a graph $G$ and an $n$-bifactor state $\rho$. The graph $G = (V,E)$ has a set of vertices $V$ and a set of edges $E$. Each $v \in V$ is a quantum system, with Hilbert space $\cH_v$. A $n$-bifactor state $\rho$ is a positive operator on $\cH = \bigotimes_{v \in V} \cH_v$, that can be expressed as
\begin{equation}
\rho = \frac{1}{Z} \Big(\prod_{v \in V} \mu_v \Big) \sn  \Big(\bigodot_{(u,v) \in E} \nu_{u:v} \Big) 
\label{eq:BS}
\end{equation}
 where $Z$ is some normalization factor, and $\mu_v$ and $\nu_{u:v}$ are positive operators on $\cH_u$ and $\cH_u \otimes \cH_v$, respectively. The operators $\nu_{u:v}$ are required to mutually commute when $n$ is finite. The product $\sn$ is defined as $X\sn Y \equiv [X^{\frac{1}{2n}}Y^{\frac 1n} X^{\frac{1}{2n}}]^n$, and has the property of producing a positive operator when both $X$ and $Y$ are positive. This product is non-commutative except in the limit $n\rightarrow \infty$, which defines the $\odot$ product: $X \odot Y \equiv \lim_{n\rightarrow \infty} X\sn Y =  e^{(\log X+\log Y)}$. Both products $\sn$ and $\odot$ reduce to normal matrix product when $X$ and $Y$ commute. 

A generic inference problem on a graphical model is to compute the reduced density operator on a subset $W \subset V$ of the quantum systems conditioned on the fact that a measurement was performed on a disjoint subset $U \subset V$, where both $W$ and $U$ are of constant size. This problem turns out to be equivalent to the seemingly simpler problem of computing the reduced state on any subset $W$ of constant size, i.e.  $\rho_{W} = \Tr_{V-W}\{\rho\}$, where $\Tr_X$ denotes the partial trace over a systems in set $X$. Without additional assumptions on the structure of $\rho$, solving this problem requires resources that grow exponentially with the number of quantum systems $|V|$. However, the solution can sometimes be obtained or approximated by QBP in a time polynomial in $|V|$. 

\section{QBP algorithm}
 Given a graphical model $(G,\rho)$, QBP consists of a sequence of exchanges of operator-valued messages between neighboring vertices, which carry information about the state at other locations in the graph. More precisely, for $(u,v) \in E$, the message passed from vertex $u$ to vertex $v$ at time $t$ is an operator on $\cH_v$ given by
\begin{equation}
m_{u\rightarrow v}(t) \propto \Tr_{u}\Big\{\mu_u \sn \Big( \nu_{u:v} \odot\!\!\!\!\!\! \bigodot_{v' \in n(u)-v} \!\!\!\!\!\! m_{v'\rightarrow u}(t-1)\Big) \Big\}
\label{eq:BP}
\end{equation}
where $n(u)$ denotes the neighbors of $u$. The proportionality factor can be chosen so that $\Tr\{m_{u\rightarrow v}\} =1$, and the messages are initialized $m_{u\rightarrow v}(0) = I$. At time $t$ the belief $b_{uv}(t)$ --- which is meant to represent some approximation of the state $\rho_{uv} = \Tr_{V-uv}\{\rho\}$ for $(u,v) \in E$ --- is given by
\begin{equation}
b_{uv}(t) \propto (\mu_u\mu_v)\sn \Big( \nu_{u:v} \odot\!\!\!\!\! \bigodot_{w\in n(u)-v}\!\!\!\!\! m_{w\rightarrow u} \!\!\!\! \bigodot_{y\in n(v)-u}\!\!\!\!\! m_{y\rightarrow v}\Big)
\label{eq:belief}
\end{equation}
where all messages are taken at time $t$. When all operators defining the bifactor state commute, QBP reduces to the standard belief propagation algorithm \cite{AM00a,Yed01a,Mac03a,MM07a}. 

Since the message update rule Eq.~\eqref{eq:BP} at vertex $u$ depends only on the incoming messages at that vertex, the algorithm can be operated in a highly parallel fashion where each quantum system $u$ is associated with a processor, and messages are exchanged between processors $u$ and $v$ iff $(u,v) \in E$. Similarly, the beliefs on the pair $(u,v)$, \Eq{belief}, can be computed by combining the messages received at those vertices.  

\subsection{Convergence}

In \cite{LP07a}, it was shown that when $G$ is a tree and $(G,\rho)$ is either {\em i}) a 1-bifactor state [c.f. Eq.~\eqref{eq:BS} with $n=1$] or {\em ii}) a quantum Markov network, 
QBP yields the exact solution in a time proportional to the graph's diameter --- i.e. $b_{uv}(t) = \rho_{uv}$ for $t\geq diameter(G)$. Intuitively, this means that the algorithm must run for a time sufficiently long to allow messages to travel between any pair of vertices.  When operated on loopy graphs, the beliefs do not necessarily converge to the correct density operators. A good heuristic in that case is to halt the algorithm when $b_{uv}(t)$ become almost time-independent, which also happens in a time roughly equal to the graph's diameter in all the models we have investigated.

A graphical model $(G,\rho)$ is a quantum Markov network when the conditional independence conditions $I(U:(V-n(U)-U)|n(U)) = 0$ are met for all $U \subset V$. The quantity $I(A:B|C) = S(AC)+S(BC) - S(C) - S(ABC)$ is the quantum conditional mutual information \cite{LR73a,HJPW03a}, and $S(A) = \Tr\{\rho_A \log_2 \rho_A\}$ is the von Neumann entropy. As explained in \cite{Rus02b,LP07a}, the vanishing of $I(A:C|B)$ is equivalent to the condition $\rho_{ABC} = \rho_B^{-1} \odot \rho_{AB} \odot \rho_{BC}$. This equality is not verified in general, and the Kullback-Leibler distance between the right- and left-hand side is precisely the conditional mutual information $D(\rho_{ABC} || \rho_B^{-1} \odot \rho_{AB} \odot \rho_{BC}) = I(A:C|B)$.

To understand the workings of QBP, consider a bifactor state $\rho_{uvw}$ on the line $u-v-w$. The reduced state on $w$ is $\rho_w \propto  \Tr_{uv}\big\{(\mu_u \mu_v \mu_w) \sn (\nu_{u:v} \odot \nu_{v:w}) \big\} $. When $n=1$, basic algebra implies that $\rho_w \propto \Tr_{v}\big\{(\mu_v \mu_w) \star^{(1)} \big(\Tr_u\{\mu_u \star^{(1)} \nu_{u:v}\} \odot\nu_{v:w}\big) \big\} $; the operations $\Tr_u$ and $\star^{(1)}$ commute so to say. The computation of $\rho_w$ can thus be broken into two steps: {\em i}) Compute $m_{u\rightarrow v} = \Tr_u\{\mu_u \star^{(1)} \nu_{u:v}\}$; {\em ii}) Compute $\rho_w \propto \Tr_{v}\big\{(\mu_v \mu_w) \star^{(1)} \big(m_{u\rightarrow v} \odot\nu_{v:w}\big) \big\}$. When $n\rightarrow \infty$ on the other hand, the operations $\Tr_u$ and $\odot$ do not commute in general, but they do precisely when $I(u:w|v) = 0$ \cite{LP07a}. QBP is based on a generalization of these observations to arbitrary graphs. 

QBP does not rely on the vanishing of the normalized  connected correlation functions $\cC(\sigma_A,\sigma_C) = \langle \sigma_A\sigma_C\rangle - \langle \sigma_A\rangle\langle \sigma_C \rangle$ \cite{WVHC07a}, or equivalently \cite{Fan73a,FvdG99a} on the vanishing of  the mutual information $I(A:C) = S(A)+S(C) -S(AC)$. In many systems, the mutual information is not {\em a priori} short range. For instance in the $T \rightarrow 0$ limit, the thermal state of the 1-d Ising model in zero transverse field is an equal mixture of all spins up and all spins down, which has $I(A:B) = 1$ between any two disjoint regions, whereas $I(A:C|B) = 0$ for any three disjoint regions. To compute thermodynamical quantities, one generally introduces a symmetry-breaking field that randomly singles out either the all-up or all-down state, which both have $I(A:B) = 0$. Symmetry-breaking can be a delicate issue --- for instance, on Cayley trees where a constant fraction of vertices live on the boundary \cite{MP01a} --- and is circumvented by QBP. 

\section{QBP for quantum many-body}

In the context of quantum many-body physics, the inference problem consists of computing correlation functions for the thermal state of a system of interacting particles. Given a graph $G = (V,E)$, we consider the generic Hamiltonian
\begin{equation}
H = \sum_{v \in V} h_v + \sum_{(u,v) \in E} h_{uv}.
\label{eq:hamiltonian}
\end{equation}
The thermal state at inverse temperature $\beta = 1/T$ is given by $\rho = \frac 1Z e^{-\beta H}$ where $Z = \Tr\{e^{-\beta H}\}$ is the partition function. Defining $\mu_v = e^{-\beta h_v}$ and $\nu_{u:v} = e^{-\beta h_{uv}}$ enables us to express any such thermal state as an $\infty$-bifactor state, cf. Eq.~\eqref{eq:BS}. 

Despite the fact that thermal states are bifactor states, the result from \cite{LP07a} cited above does not imply that correlation functions can be evaluated exactly and efficiently with QBP. This is primarily because $G$ is not necessarily a tree, but also because thermal states are neither 1-bifactor  nor quantum Markov networks in general. There is no general remedy to the first hurdle, unless the loops happen to be very small and can be eliminated by merging some vertices. Thus, QBP will need to be executed on a loopy graph and it is the primary goal of this paper to determine the effects of such loops on the performance of QBP. Two pragmatic solutions, named the replica method and sliding window QBP, have been proposed to overcome the second set of obstacles \cite{LP07a}. 

\subsection{Replica} 

The general idea of the replica method is to approximate the thermal state by a 1-bifactor state on which QBP can be executed directly and is guaranteed to converge in the absence of loops. In a first step, a Trotter-Suzuki (TS) decomposition is used to approximate a thermal state by an $N_\tau$-bifactor state with finite $N_\tau$. This produces a systematic error that scales as $\beta/N_\tau$. Then, in a fashion reminiscent of the replica trick used in the study of spin glass, the $N_\tau$-bifactor state is replaced by a 1-bifactor state at the expense of substituting the quantum system at each vertex by $N_\tau$ replicas: 
\begin{equation}
\mu_v \rightarrow \Big(\mu_v^{\frac{1}{N_\tau}}\Big)^{\otimes N_\tau} T_v^{(N_\tau)} \ \ {\rm and} \ \
\nu_{u:v} \rightarrow \Big(\nu_{u:v}^{\frac{1}{N_\tau}}\Big)^{\otimes N_\tau}
\end{equation}
where $T_v^{(N_\tau)}$ is the operator that cyclicly permutes the $N_\tau$ replicas of $v$. The operators $\nu_{u:v} = e^{-\beta h_{uv}}$ do not commute in general, but this can be fixed in practice on sparse graphs by merging some vertices. On a tree, the TS decomposition is the only source of error, so  accuracy $\epsilon$ can be achieved at a computational cost that is exponential in $\beta/\epsilon$. This method is particularly useful as it allows for a direct computation of correlation functions at arbitrary distances, see \cite{LP07a}.

\subsection{Sliding window}

While all quantum Markov networks are thermal states of some local Hamiltonian on $G$ \cite{LP07a}, the converse is not true in general. Sliding window QBP is motivated by the fact that quantum Markov networks are fixed points of coarse graining procedures.  Thermal states, regarded as $\infty$-bifactor states, are used directly to implement the message passing rule in Eq.~\eqref{eq:BP} with $n = \infty$, except that messages are computed not just using the nearest neighbors but with all vertices within a distance $\leq \ell$. On a line, for instance, vertex $j$ receives a message from $(j-1,j-2,\ldots j-\ell)$ and one from $(j+1,j+2,\ldots,j+\ell)$. In that case, sliding window QBP produces the exact solution efficiently if the conditional mutual information dies off at a finite distance.

\section{Numerical results}

We have numerically implemented the QBP algorithm on various graphs for the Ising and Heisenberg model whose Hamiltonians are
\begin{equation}
H_{I} = \sum_{v \in V} \vec g\cdot \vec \sigma_v + \!\!\sum_{(u,v) \in E} \!\! J_{uv}\sigma^z_u\sigma^z_v \ \ \mathrm{and}\ \  H_{H} = \!\!\! \sum_{(u,v) \in E} \!\!\! \vec \sigma_u \cdot \vec\sigma_v
\end{equation}
respectively, and $\vec\sigma = (\sigma^x,\sigma^y,\sigma^x)$ are the usual Pauli matrices normalized so that $\sigma^2 = \II /4$. On a line, the homogeneous ($J_{uv} = 1$) Ising model has a zero temperature phase transition at the critical transverse field $\vec g=(\frac 12,0,0)$. Most of our simulations were performed at this critical value, as it is expected to represent the ``hardest case". Unless otherwise specified, it is henceforth assumed that $\vec g=(\frac 12,0,0)$ and $J_{uv} = 1$. 

\begin{figure}
\center\includegraphics[width=7.5cm]{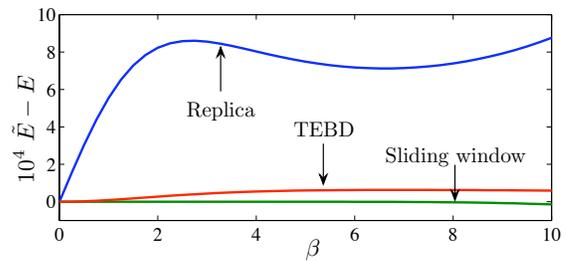}
\caption{(Color online) Critical Ising model on infinite line. Energy density estimate $\tilde E$ using the method of replicas with $N_\tau = 10$, sliding window QBP with $\ell = 6$, and TEBD with $\chi = 150$, compared to the exact energy density $E$ obtained from fermionization. }
\label{fig:line}
\end{figure}

We used QBP to compute the energy density of the Ising model on an infinite line. This model can be solved exactly by means of a Jordan-Wigner transform that maps the interacting spin chain to a collection of free fermions \cite{Pfe70a}.  Figure~\ref{fig:line} shows the difference between the energy density computed with QBP and its exact value as a function of inverse temperature. Also shown are results obtained from a superoperator version of time-evolving block decimation (TEBD), which combines ideas from \cite{Vid06a,ZV04a}. Since a line is a tree, the error in the results obtained from the replica method is entirely caused by the TS decomposition. The results obtained for sliding window QBP are in remarkably good agreement with the exact value, and can be systematically improved by increasing $\ell$. This reflects the fact that correlations are short-ranged in finite-temperature 1D models. As expected the agreement improves for non-critical $g$. Results obtained for the Heisenberg model on the infinite line (not shown)  are similar in all aspects. 

\begin{figure}
\center\includegraphics[width=7.5cm]{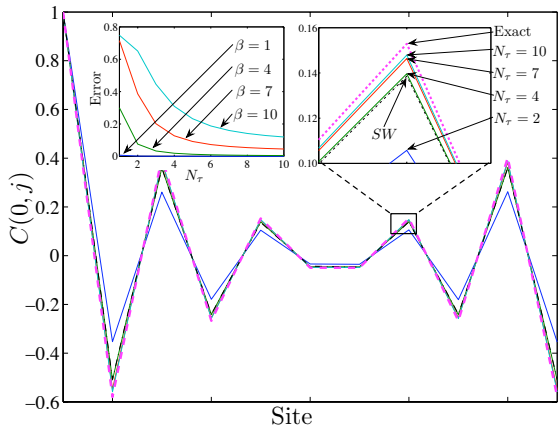}
\caption{(Color online) Correlations for $H_I$ on a 11-site circle at $\beta = 6$. Exact numerical solution (dash), sliding window with $\ell = 5$ (dash-dot), and the replica method for various values of $N_\tau$ (full). Left inset: Error Eq.~\eqref{eq:error} {\em vs} the $N_\tau$ for different $\beta$.}
\label{fig:corr}
\end{figure}

To characterize the performance of QBP on more general graphs, we restrict our attention to systems with less than 12 spins, allowing comparison to direct brute-force numerical solutions. Figure \ref{fig:corr} shows the correlation function $C(0,j) = \Tr\{\sigma^z_0\sigma^z_j \rho \}$ for $H_I$ on a frustrated 11-site circle. We assess the quality of the approximation $\tilde C$ to the exact correlation $C$ by the average relative error
\begin{equation}
{\rm Error} = \frac{\sum_j |C(0,j) - \tilde C(0,j)|}{\sum_j |C(0,j)|}.
\label{eq:error}
\end{equation}
Sliding window is again in very good agreement with the exact value for a relatively small window size. For the values of $N_\tau \leq 10$ accessible with modest computational resources, the replica QBP reproduces the exact correlation function within a few percents at sufficiently high temperatures $\beta \lesssim 6$, which is consistent with the systematic error due to the TS decomposition.

\begin{figure}[!tbh]
\center\includegraphics[width=8.6cm]{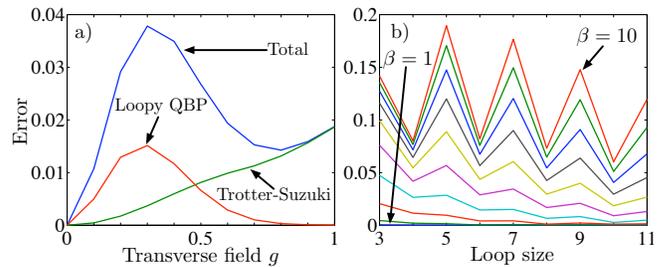}
\caption{(Color online) Replica method with $N_\tau = 10$. a) Different contributions to the error Eq.~\eqref{eq:error} {\em vs} transverse field for $H_I$ on a 11-sites circle at $\beta = 6$. b) Error {\em vs} loop size, for $\beta = 1,2,\ldots 10$. }
\label{fig:loop}
\end{figure}

Indeed, both the TS decomposition and the loopy QBP contribute to the total error Eq.~\eqref{eq:error}. By brute force computation, it is possible to determine exactly what fraction of the error is caused by each of these approximations, and in almost all cases we have studied at critical $g$, both contributions were comparable. Figure \ref{fig:loop}a shows each contribution to the total error as a function of the transverse field $\vec g = (g,0,0)$. 

The most successful applications of classical belief propagation algorithm are on graphs whose typical loop size is very large. This is the case for instance of low density parity check codes \cite{Gal63a,RU05a} and spin glasses on Bethe lattices \cite{MP01a}. Intuitively, one expects a local algorithm like belief propagation to be relatively insensitive to the large-scale structure of the graph. We expect QBP to share this feature, and Figure \ref{fig:loop}b illustrates the effect of the loop size on the average relative error of the correlation function. The oscillatory behavior of the error is explained by the frustration present in odd-size circles. Save from these oscillations, the results show a global improvement as the loop size increases. Errors obtained from sliding window (not shown) also show a clear improvement as the loop size increases, but tend to have higher errors on even-size loops.

\begin{figure}
\center\includegraphics[width=7.5cm]{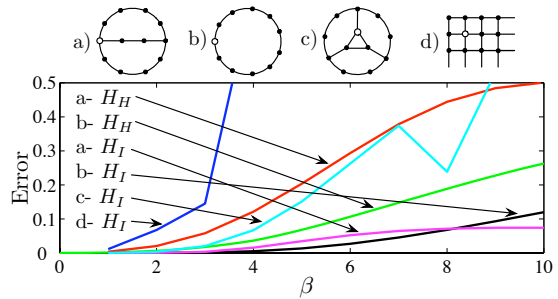}
\caption{(Color online) Error Eq.~\eqref{eq:error} for Ising and Heisenberg models on various loopy graphs using the replica method. For the Ising model $N_\tau = 10$ and for the Heisenberg model $N_\tau = 4$. [d) is a torus.]}
\label{fig:graph}
\end{figure}

We have tested QBP on a variety of graphs depicted on Fig.~\ref{fig:graph} a)-d). The resulting errors in the correlation functions are shown in Fig.~\ref{fig:graph}. The computational cost is slightly higher for the Heisenberg model because $h_{uv}$ do not mutually commute. This restricts the computation to lower values of $N_\tau$ and consequently yields larger errors. Modulo this difference, the error is most prominent for graphs c) and d) which contain loops of size 3. In those cases, we found that the QBP algorithm was not converging: the magnitude of the errors is consistent with the magnitude of the time fluctuations of $b_{uv}(t)$ , c.f. Eq.~\eqref{eq:belief}. As expected, the predicted correlation function is in much better agreement with its exact value on graphs a) and b) that have only relatively large loops.  

\section{Conclusion}

We have numerically characterized the performance of the recently proposed QBP algorithm. In the high temperature phase, both the replica and the sliding window QBP algorithms perform remarkably well on a tree with modest computational resources, c.f. Fig.~\ref{fig:line}, and offer performances similar to TEBD. On loopy graphs, we found that the algorithm gives reliable approximations when the loop size is large. Most importantly, when the results deviated from the exact value, e.g. in the presence of small loops, the algorithm did not reach a steady state, i.e.  the beliefs Eq.~\eqref{eq:belief} were highly fluctuating as a function of time. This provides an indirect way of assessing the validity of the results. 

In \cite{LSS07a}, a technique similar to what we have called the replica method was used to investigate the phase diagram of quantum spin-glasses on Cayley trees. Based on the results we have presented, QBP should be suitable to study this phase diagram for more general Bethe lattices whose typical loop size scales as $\log|V|$. In the classical setting, it has been argued that the physics of random Bethe lattices and Cayley trees is greatly different \cite{MP01a}. We note that the randomness in quenched disordered systems should not affect the performences of QBP. In fact, our results obtained for random couplings $J_{uv}$ and random local fields $\vec g$ are typically in better agreement than the ones we have presented.

Finally, the low temperature phase of these models may be accessible using QBP as part of a variational approach based on projected entangled-pair states \cite{VC04a}, which are a form of 1-bifactor states. QBP can be used to approximately compute the reduced state on pairs of sites and minimize their energy. We leave the characterization of this approach for a future study. 

\medskip 
\noindent{\bf{\sf Acknowledgments}} --- 
We thank Matt Leifer for stimulating discussions on graphical models and QBP. DP is supported in part by the Gordon and Betty Moore Foundation, by the NSF under Grants No. PHY-0456720, and by NSERC.


\begin{thebibliography}{23}
\expandafter\ifx\csname natexlab\endcsname\relax\def\natexlab#1{#1}\fi
\expandafter\ifx\csname bibnamefont\endcsname\relax
  \def\bibnamefont#1{#1}\fi
\expandafter\ifx\csname bibfnamefont\endcsname\relax
  \def\bibfnamefont#1{#1}\fi
\expandafter\ifx\csname citenamefont\endcsname\relax
  \def\citenamefont#1{#1}\fi
\expandafter\ifx\csname url\endcsname\relax
  \def\url#1{\texttt{#1}}\fi
\expandafter\ifx\csname urlprefix\endcsname\relax\def\urlprefix{URL }\fi
\providecommand{\bibinfo}[2]{#2}
\providecommand{\eprint}[2][]{\url{#2}}

\bibitem[{\citenamefont{Aji and Mc{E}liece}(2000)}]{AM00a}
\bibinfo{author}{\bibfnamefont{S.}~\bibnamefont{Aji}} \bibnamefont{and}
  \bibinfo{author}{\bibfnamefont{R.}~\bibnamefont{Mc{E}liece}},
  \bibinfo{journal}{IEEE Trans. Info. Theor.} \textbf{\bibinfo{volume}{46}},
  \bibinfo{pages}{325} (\bibinfo{year}{2000}).

\bibitem[{\citenamefont{MacKay}(2003)}]{Mac03a}
\bibinfo{author}{\bibfnamefont{D.~J.~C.} \bibnamefont{MacKay}},
  \emph{\bibinfo{title}{Information Theory, Inference and Learning Algorithms}}
  (\bibinfo{publisher}{Cambridge University Press},
  \bibinfo{address}{Cambridge, UK}, \bibinfo{year}{2003}).

\bibitem[{\citenamefont{M\'ezard and Montanari}(2007)}]{MM07a}
\bibinfo{author}{\bibfnamefont{M.}~\bibnamefont{M\'ezard}} \bibnamefont{and}
  \bibinfo{author}{\bibfnamefont{A.}~\bibnamefont{Montanari}},
  \emph{\bibinfo{title}{Constraint Satisfaction Networks in Physics and
  Computation}} (\bibinfo{publisher}{Clarendon Press}, \bibinfo{year}{2007}).

\bibitem[{\citenamefont{Yedidia}(2001)}]{Yed01a}
\bibinfo{author}{\bibfnamefont{J.~S.} \bibnamefont{Yedidia}},
  \emph{\bibinfo{title}{Advanced mean field methods: theory and practice}}
  (\bibinfo{publisher}{MIT Press}, \bibinfo{year}{2001}), chap.
  \bibinfo{chapter}{An idiosyncratic journey beyond mean field theory},
  p.~\bibinfo{pages}{21}.

\bibitem[{\citenamefont{Berrou et~al.}(1993)\citenamefont{Berrou, Glavieux, and
  Thitimajshima}}]{BGT93a}
\bibinfo{author}{\bibfnamefont{C.}~\bibnamefont{Berrou}},
  \bibinfo{author}{\bibfnamefont{A.}~\bibnamefont{Glavieux}}, \bibnamefont{and}
  \bibinfo{author}{\bibfnamefont{P.}~\bibnamefont{Thitimajshima}}, in
  \emph{\bibinfo{booktitle}{ICC'93}} (\bibinfo{address}{Gen\`eve, Switzerland},
  \bibinfo{year}{1993}), pp. \bibinfo{pages}{1064--1070}.

\bibitem[{\citenamefont{Gallager}(1963)}]{Gal63a}
\bibinfo{author}{\bibfnamefont{R.~G.} \bibnamefont{Gallager}},
  \emph{\bibinfo{title}{Low Density Parity Check Codes}}
  (\bibinfo{publisher}{M.I.T. Press}, \bibinfo{address}{Cambridge,
  Massachusetts}, \bibinfo{year}{1963}).

\bibitem[{\citenamefont{Richardson and Urbanke}(2008)}]{RU05a}
\bibinfo{author}{\bibfnamefont{T.}~\bibnamefont{Richardson}} \bibnamefont{and}
  \bibinfo{author}{\bibfnamefont{R.}~\bibnamefont{Urbanke}},
  \emph{\bibinfo{title}{Modern Coding Theory}} 
   (\bibinfo{publisher}{Cambridge University Press},
  \bibinfo{address}{Cambridge, UK}, \bibinfo{year}{2008}).

\bibitem[{\citenamefont{M\'ezard and Parisi}(2001)}]{MP01a}
\bibinfo{author}{\bibfnamefont{M.}~\bibnamefont{M\'ezard}} \bibnamefont{and}
  \bibinfo{author}{\bibfnamefont{G.}~\bibnamefont{Parisi}},
  \bibinfo{journal}{The European Physical Journal B}
  \textbf{\bibinfo{volume}{20}}, \bibinfo{pages}{217} (\bibinfo{year}{2001}).
  
\bibitem[{\citenamefont{M\'ezard et~al.}(2002)\citenamefont{M\'ezard, Parisi,
  and Zecchina}}]{MPZ02a}
\bibinfo{author}{\bibfnamefont{M.}~\bibnamefont{M\'ezard}},
  \bibinfo{author}{\bibfnamefont{G.}~\bibnamefont{Parisi}}, \bibnamefont{and}
  \bibinfo{author}{\bibfnamefont{R.}~\bibnamefont{Zecchina}},
  \bibinfo{journal}{Science} \textbf{\bibinfo{volume}{297}},
  \bibinfo{pages}{812} (\bibinfo{year}{2002}).

\bibitem[{\citenamefont{Monasson et~al.}(1999)\citenamefont{Monasson, Zecchina,
  Kirkpatrick, Selman, and Troyansky}}]{MZK+99a}
\bibinfo{author}{\bibfnamefont{R.}~\bibnamefont{Monasson}},
  \bibinfo{author}{\bibfnamefont{R.}~\bibnamefont{Zecchina}},
  \bibinfo{author}{\bibfnamefont{S.}~\bibnamefont{Kirkpatrick}},
  \bibinfo{author}{\bibfnamefont{B.}~\bibnamefont{Selman}}, \bibnamefont{and}
  \bibinfo{author}{\bibfnamefont{L.}~\bibnamefont{Troyansky}},
  \bibinfo{journal}{Nature} \textbf{\bibinfo{volume}{400}},
  \bibinfo{pages}{133} (\bibinfo{year}{1999}).

\bibitem[{\citenamefont{Hastings}(2007)}]{Has07b}
\bibinfo{author}{\bibfnamefont{M.B.}~\bibnamefont{Hastings}},
  \bibinfo{journal}{Phys. Rev. B} \textbf{\bibinfo{volume}{76}},
  \bibinfo{pages}{201102(R)} (\bibinfo{year}{2007}).

\bibitem[{\citenamefont{Leifer and Poulin}(2007)}]{LP07a}
\bibinfo{author}{\bibfnamefont{M.}~\bibnamefont{Leifer}} \bibnamefont{and}
  \bibinfo{author}{\bibfnamefont{D.}~\bibnamefont{Poulin}},
  \bibinfo{journal}{Ann. Phys., in press}.

\bibitem[{\citenamefont{Hayden et~al.}(2004)\citenamefont{Hayden, Jozsa, Petz,
  and Winter}}]{HJPW03a}
\bibinfo{author}{\bibfnamefont{P.}~\bibnamefont{Hayden}},
  \bibinfo{author}{\bibfnamefont{R.}~\bibnamefont{Jozsa}},
  \bibinfo{author}{\bibfnamefont{D.}~\bibnamefont{Petz}}, \bibnamefont{and}
  \bibinfo{author}{\bibfnamefont{A.}~\bibnamefont{Winter}},
  \bibinfo{journal}{Comm. Math. Phys.} \textbf{\bibinfo{volume}{246}},
  \bibinfo{pages}{359} (\bibinfo{year}{2004}).

\bibitem[{\citenamefont{Lieb and Ruskai}(1973)}]{LR73a}
\bibinfo{author}{\bibfnamefont{E.}~\bibnamefont{Lieb}} \bibnamefont{and}
  \bibinfo{author}{\bibfnamefont{M.}~\bibnamefont{Ruskai}},
  \bibinfo{journal}{J. Math. Phys.} \textbf{\bibinfo{volume}{14}},
  \bibinfo{pages}{1938} (\bibinfo{year}{1973}).

\bibitem[{\citenamefont{Ruskai}(2002)}]{Rus02b}
\bibinfo{author}{\bibfnamefont{M.~B.} \bibnamefont{Ruskai}},
  \bibinfo{journal}{J. Math. Phys.} \textbf{\bibinfo{volume}{43}},
  \bibinfo{pages}{4358} (\bibinfo{year}{2002}).

\bibitem[{\citenamefont{Wolf et~al.}(2007)\citenamefont{Wolf, Verstraete,
  Hastings, and Cirac}}]{WVHC07a}
\bibinfo{author}{\bibfnamefont{M.~M.} \bibnamefont{Wolf}},
  \bibinfo{author}{\bibfnamefont{F.}~\bibnamefont{Verstraete}},
  \bibinfo{author}{\bibfnamefont{M.~B.} \bibnamefont{Hastings}},
  \bibnamefont{and} \bibinfo{author}{\bibfnamefont{J.~I.} \bibnamefont{Cirac}},
  (\bibinfo{year}{2007}), \eprint{arXiv.org:0704.3906}.

\bibitem[{\citenamefont{Fannes}(1973)}]{Fan73a}
\bibinfo{author}{\bibfnamefont{M.}~\bibnamefont{Fannes}},
  \bibinfo{journal}{Comm. Math. Phys.} \textbf{\bibinfo{volume}{31}},
  \bibinfo{pages}{291} (\bibinfo{year}{1973}).

\bibitem[{\citenamefont{Fuchs and van~de Graaf}(1999)}]{FvdG99a}
\bibinfo{author}{\bibfnamefont{C.~A.} \bibnamefont{Fuchs}} \bibnamefont{and}
  \bibinfo{author}{\bibfnamefont{J.}~\bibnamefont{van~de Graaf}},
  \bibinfo{journal}{IEEE Trans. Info. Theor.} \textbf{\bibinfo{volume}{45}},
  \bibinfo{pages}{1216} (\bibinfo{year}{1999}).

\bibitem[{\citenamefont{Pfeuty}(1970)}]{Pfe70a}
\bibinfo{author}{\bibfnamefont{P.}~\bibnamefont{Pfeuty}},
  \bibinfo{journal}{Ann. of Phys.} \textbf{\bibinfo{volume}{57}},
  \bibinfo{pages}{79} (\bibinfo{year}{1970}).

\bibitem[{\citenamefont{Vidal}(2006)}]{Vid06a}
\bibinfo{author}{\bibfnamefont{G.}~\bibnamefont{Vidal}},
  \bibinfo{journal}{Phys. Rev. Lett.} \textbf{\bibinfo{volume}{98}},
  \bibinfo{pages}{070201} (\bibinfo{year}{2007}).

\bibitem[{\citenamefont{Zwolak and Vidal}(2004)}]{ZV04a}
\bibinfo{author}{\bibfnamefont{M.}~\bibnamefont{Zwolak}} \bibnamefont{and}
  \bibinfo{author}{\bibfnamefont{G.}~\bibnamefont{Vidal}},
  \bibinfo{journal}{Phys. Rev. Lett.} \textbf{\bibinfo{volume}{93}},
  \bibinfo{pages}{207205} (\bibinfo{year}{2004}).

\bibitem[{\citenamefont{Laumann et~al.}(2007)\citenamefont{Laumann,
  Scardicchio, and Sondhi}}]{LSS07a}
\bibinfo{author}{\bibfnamefont{C.}~\bibnamefont{Laumann}},
  \bibinfo{author}{\bibfnamefont{A.}~\bibnamefont{Scardicchio}},
  \bibnamefont{and} \bibinfo{author}{\bibfnamefont{S.}~\bibnamefont{Sondhi}},
 (\bibinfo{year}{2007}), \eprint{arXiv:0706.4391}.

\bibitem[{\citenamefont{Verstraete and Cirac}(2004)}]{VC04a}
\bibinfo{author}{\bibfnamefont{F.}~\bibnamefont{Verstraete}} \bibnamefont{and}
  \bibinfo{author}{\bibfnamefont{J.~I.} \bibnamefont{Cirac}},
 (\bibinfo{year}{2004}),  \eprint{cond-mat/0407066}.

\end{thebibliography}

\end{document}